# Full-resolution Lung Nodule Segmentation from Chest X-ray Images using Residual Encoder-Decoder Networks

Michael James Horry [a,b], Subrata Chakraborty [c,*], Biswajeet Pradhan [a,d], Manoranjan Paul [e], Jing Zhu [f], Prabal Datta Barua [c,g,h], U. Rajendra Acharya [i], Fang Chen [j], Jianlong Zhou [j]

[a] Centre for Advanced Modelling and Geospatial Information Systems (CAMGIS), School of Civil and Environmental Engineering, Faculty of Engineering and IT, University of Technology Sydney, Ultimo, NSW 2007, Australia

[b] IBM Australia Limited, Sydney, NSW 2000, Australia

[c] School of Science and Technology, Faculty of Science, Agriculture, Business and Law, University of New England, Armidale, NSW 2351, Australia

[d] Earth Observation Centre, Institute of Climate Change, Universiti Kebangsaan Malaysia, 43600 UKM, Bangi, Selangor, Malaysia

[e] Machine Vision and Digital Health (MaViDH), School of Computing, Mathematics, and Engineering, Charles Sturt University, Bathurst, NSW 2795, Australia

[f] Department of Radiology, Westmead Hospital, Westmead, NSW 2145, Australia

[g] Cogninet Brain Team, Cogninet Australia, Sydney, NSW 2010, Australia

[h] School of Business (Information Systems), Faculty of Business, Education, Law & Arts, University of Southern Queensland, Toowoomba, QLD 4350, Australia

[i] School of Mathematics, Physics and Computing, University of Southern Queensland, Springfield, QLD 4300, Australia

[j] Data Science Institute, University of Technology Sydney, Ultimo, NSW 2007

*   Correspondence: Subrata.Chakraborty@une.edu.au

**Abstract**

Lung cancer is the leading cause of cancer death and early diagnosis is associated with a positive prognosis. Chest X-ray provides an inexpensive imaging mode for lung cancer diagnosis, especially for remote populations with limited healthcare access. Unfortunately, suspicious nodules are difficult to distinguish from pulmonary vascular and bone structures using Chest X-ray. Computer vision algorithms have previously been proposed to assist human radiologists in this task, however, the leading studies use down-sampled images and highly parameterized, computationally expensive methods with unproven generalization capability to unseen datasets. Instead, this study localizes lung nodules from chest X-ray images using highly efficient encoder-decoder neural networks that have been crafted to process full resolution input images to avoid any signal loss resulting from down-sampling.

Encoder-decoder networks are trained and tested internally using the Japanese Society of Radiological Technology lung nodule dataset. The trained networks are used to localize lung nodules from an independent external chest x-ray dataset. Sensitivity and false positive rates are measured using an automated framework to eliminate any observer subjectivity. These experiments allow for the determination of the optimal network depth, image resolution and pre-processing pipeline for generalized lung nodule localization from chest x-ray images.

We find that nodule localization is influenced by subtlety, with more subtle nodules being detected in earlier training epochs. Therefore, we propose a novel self-ensemble model from three consecutive epochs centered on the conventional validation optimum. This ensemble achieved a sensitivity of 85% in 10-fold internal testing with false positives of 8 per image. A sensitivity of 81% is achieved at a false positive rate of 6 following morphological false positive reduction. This result is comparable to more computationally complex systems based on linear and spatial filtering, but with a sub-second inference time that is an order of magnitude faster than other methods presented in the literature. The proposed algorithm achieved excellent generalization results against a challenging external dataset with a sensitivity of 77% at a false positive rate of 7.6, thereby proving robustness.

# 1 Introduction

Lung cancer is the third most common cancer globally, and it is the most deadly accounting for around 1 in 5 cancer deaths [1]. In the United States, lung cancer deaths have declined by around 50% since 1990 for males and by 30% from 2002 for females; this reduction is attributable to reductions in smoking and improvements in early detection and treatment [2]. Unfortunately, the United States' pattern of lung cancer decline is not global. Whilst lung cancer has declined for males in 19 countries, it has increased for women in 26 countries over the same period. Concerningly, although lung cancer is associated with aging populations [3], the risk among younger generations shows increasing trends in some geographies [4].

This paper focusses on the automatic localization of lung nodules, small roughly spherical growths on the lung, from Chest X-ray (CXR) images. The detection of lung nodules is important because they often represent early-stage lung cancer [5], and there is a directly proportional relationship between the size and growth rate of a nodule, and nodule malignancy [6]. On a typical workday, a clinical radiologist may be required to analyze a CXR image every 3-4 seconds [7]. Over a work shift of 8-10 hours, it can be expected that radiologists experience fatigue, leading to potential drifting of attention, with a negative effect on the quality of diagnosis [8]. Automated tools that facilitate the tracking of lung nodules have been shown to help reduce lung cancer mortality, and associated public health costs [9], estimated to average around $51,000 per case in Australia [10] in 2020, and $45,897 per case in the US in the most recent comparable assessment in 2005 [11].

Segmentation of lung nodules from CXR images is a challenging task since the complex background of pulmonary cavity anatomy makes it difficult to distinguish a true positive lung nodule from a false positive feature such as a vascular bundle or rib crossing. The lung nodule itself appears as a relatively small feature against a large and noisy background. This differs from typical computer vision/classification tasks documented in the literature where the image feature of interest, say cat or dog, appears as a large foreground shape that dominates the overall image, as is the case for the frequently cited ImageNet archive [12].

Studies into automated lung nodule detection from CXR images usually perform feature extraction tasks using parameterized linear and spatial filtering methods such as wavelet, convergence index, sliding-band, and Laplacian of Gaussian (LoG) [13-17]. Deep-learning algorithms are then used for the classification of these features as nodule/non-nodule for false positive reduction. The filter parameters for these feature extraction methods are determined empirically and require retuning where datasets exhibit systematic quality differences [18]. Hence, the generalization properties of nodule detection systems based on LoG and other filtering techniques tend to be relatively poor, even when applied to relatively high--resolution CT-based imaging [19]. Filtering-based feature extraction can also result in long inference times due to resource-intensive pixel-wise calculations. For example, [20] reported image processing times of up to 70 seconds per image. In a clinical context, where radiologists are required to interpret an image every few seconds [7], long inferencing times will adversely impact radiologists' workflow and hinder clinical adoption.

Previous studies into nodule detection from CXR images using an end-to-end deep learning pipeline employ a sliding-window segmentation methodology. This creates a small patch around every pixel in a CXR image, using the nodule/non-nodule label for that image patch as the training data for an ensemble of CNNs. Such methods result in millions of image patches from a typical 2048 x 2048-pixel image and are resource intensive to train. The leading study employing this technique achieved near-perfect internal testing results, however, external testing resulted in an average 55% reduction in classification metrics prior to fine-tuning [21].

This study localizes lung nodules from CXR images using highly efficient encoder-decoder (E-D) neural networks. The E-D networks have been crafted to process full resolution input images to avoid signal loss resulting from down-sampling. The resulting algorithm can localize lung nodules from CXR images in sub-second inference time with high sensitivity. To achieve this, we train a series of E-D neural networks based on U-Net [22], using a training set of 140 lung nodule examples with nodule masks generated from metadata provided by [23]. The E-D network is modified in several novel ways to achieve acceptable performance in this difficult task. The first innovation is a modification of the E-D filter sizes to allow network training using full-resolution CXR images. This eliminates any need to down-sample the CXR image, with the associated undesirable effect of weakening the nodule signal against the noisy CXR background. Since the E-D network is symmetrical, the decoded output is also full-resolution and provides a good indication of the location, size, and shape of predicted nodules. Secondly, following the work of [24], we improved the network recovery from difficult CXR images by using instance normalization [25] rather than batch normalization between layers in the E-D network. Thirdly, we observed that E-D network sensitivity varied by nodule subtlety and degree of network training. This problem is solved by combining the outputs of the E-D models trained to the validation optimum, along with one epoch lower and one epoch higher, thereby reinforcing nodule signal and diffusing noise.

The study also considers the influence of commonly employed pre-processing methods of global histogram equalization and lung field segmentation, showing that these operations are needed to achieve good generalization results. The effect of E-D network depth and image resolution are also explored. The resulting E-D algorithm yields excellent results that generalize well to an external dataset.

The novelty of this work can be attributed to three main aspects. Firstly, this is the first study in the literature to effectively utilize an efficient end-to-end E-D architecture for the generalized localization of lung nodules from full-resolution CXR images. This approach sets the study apart from previous works. Secondly, the proposed algorithm is externally tested using a publicly available dataset, which showcases the importance of lung field segmentation and histogram equalization as crucial pre-processing steps to achieve generalization. This aspect adds value to the study as it allows for replication by other researchers. Thirdly, this study also conducts a comprehensive analysis of the proposed algorithm's performance in relation to nodule location, diagnosis, malignancy, and patient sex. This analysis provides insights into the strengths and weaknesses of the algorithm and contributes to a more comprehensive understanding of its abilities.

## 1.1 Semantic Segmentation Overview

Semantic segmentation is the computer vision task of assigning a class label to each pixel in an image. In the case of lung nodule segmentation from CXR images, each pixel is binary classified as either being part of a lung nodule or not. There are two commonly employed methods of performing semantic segmentation tasks, sliding-window segmentation, and E-D-based methods such as U-Net.

The sliding-window method of semantic segmentation moves a small frame over the image, one pixel at a time. The frame is then classified using a ML algorithm with that classification being applied to the center pixel of the frame. This method is computationally expensive since each pixel is the center of a distinct image crop for training. For a typical 2048 x 2048-pixel CXR, this would result in 4,194,304 image crops needed for training the segmentation engine over the full-resolution image. Where multiple resolution crops are used per [21] that number of crops is multiplied by the number of different resolution frames. Even allowing for the lowered resolution of the patch frame, training will be resource-intensive and very time-consuming for this quantity of training samples. Moreover, the sliding-window segmentation method is an inefficient feature mapping algorithm, since every sliding frame overlaps most pixels with adjacent frames, resulting in features being multiplexed throughout convolutional layers by repeated input.

U-Net was originally proposed by [22] for biomedical image segmentation using cell histology datasets. U-Net is a fully convolutional encoder-decoder (E-D) network that consists of an image down-sampling encoder path connected to an up-sampling decoder path via a bridge consisting of two 3x3 CNN layers followed by a 2x2 up-sampling layer. The encoder path learns a feature map by reducing spatial information and increasing feature information, and the decoder path then precisely locates that feature of interest using transposed up-convolutions that are concatenated with the feature map from the encoding path. The result of this process is an output segmentation map given by an input image. The U-Net architecture has been incredibly successful in several applications that are diverse from the original cell histology task, including semantic segmentation of anatomical organs such as lungs [26].

## 1.2 Related Work

The personal and public health costs of lung cancer have driven a comprehensive body of research into computer-based detection of lung cancer [9, 27, 28]. It is known that radiologists fail to detect lung cancer from CXR images in around 20% of cases [29, 30]. This is concerning because CXR is frequently used as a first-presentation diagnostic for lung cancer [31], and it is known that early detection of lung cancer dramatically improves patient outcomes [32]. The idea of using computer vision methods to assist radiologists in finding lung nodules on CXR images, thereby reducing this miss rate, is appealing and research into the use of computer vision methods to automate the detection of lung nodules has been a popular field of research from the first availability of economically priced computer systems in the 1970s and 1980s [33-36].

Early studies made use of gray-level thresholding [37] and wavelet filtering [38, 39] techniques to distinguish lung nodules from other lung field tissues and structures. These methods were frequently used in combination with difference image techniques such as dual-energy subtraction [40, 41] to reduce the intensity of bony and vascular structures in the source image [37, 42], thereby reinforcing the lung nodule signal. Although these early studies delivered promising lab-based results using small test datasets of homogenous CXR images, the thresholding and filtering methods employed are inherently sensitive to the systematic image contrast and pixel intensity differences that result from CXR acquisition parameters including radiation dose, acquisition geometry, detector performance, and post-processing [43]. Thresholding techniques do not penalize false positive, or false negative pixels which makes achieving a stable threshold extremely difficult [44]. These methods, therefore, require fine-tuning of threshold values to account for the image quality differences [45], which limits generalization to real clinical settings.

More recent studies into lung nodule detection from CXR images make use of machine learning techniques (ML) to automatically find the best-fitting algorithm for nodule classification [46]. ML algorithms are trained on example images for the classes of interest, learning to differentiate diseased tissue from normal tissue using hundreds or thousands of sample images. Over the past decade, the introduction of widely available GPU computing has allowed for the cost-effective training of multi-layer neutral networks, a process known as deep learning. Deep neural networks, especially the Convolutional Neural Network (CNN) architecture proposed by [47] and first adapted to computer vision applications by [48], have excelled at many computer vision tasks including automated medical image analysis [49]. Deep-learning CNNs automatically combine feature extraction and feature classification into a single trainable algorithm [50].

The work by [18], presents the first study to use an artificial neural network (ANN) for both nodule feature extraction and classification. The authors noted that ANNs have improved generalization properties, stemming from their ability to learn from example data, eliminating the need for highly tuned *a priori* rules. Their study used a source dataset of 60 Posterior-Anterior (PA) CXR images, with each displaying single or multiple lung nodules of size 8 – 20mm, along with 60 images containing simulated small nodule images. The results achieved 94% sensitivity at 5 false positives per image for real image examples, and sensitivity of 89% at 7 false positives per image for smaller simulated test cases thereby proving the potential of using a fully ANN-based pipeline for computer vision-based lung nodule detection, albeit against a small non-standard dataset with simulated disease positive cases.

The use of a standardized lung nodule database in the automated detection of lung nodules using an ANN was first described by [15]. This study used a multiscale ANN that identified candidate nodules by multiscale blob detection leveraging Gabor wavelet and LoG filters. This candidate nodule feature space was fed into a simple ANN to classify blobs as either nodule or non-nodule. This system was trained and tested against the standard JSRT dataset [23] using a subset of 199 images with nodules in the range of 5-20mm. Under internal testing, this system achieved a sensitivity of 75% with 10.2 false positives per image. The authors of this study also took the critical action of external validation against a private dataset comprising 40 images with at least one nodule, achieving a sensitivity of 75% at a false positive rate of 8.9. This proved that ANN-based feature classification generalized well to an external dataset.

The problem of detection of very subtle lung nodules was addressed by [51] using thresholding based on edge-gradient values followed by average radial gradient filtering to create a 35-dimensional feature space which was then classified using an ANN. Experiments were performed against a private dataset consisting of 1000 CXR images with at least one lung nodule of which 73 were considered to be subtle. The subtle lung nodules of interest were in opaque areas of the CXR image such as the mediastinum, cardiac and lung area below the diaphragm. The system achieved a sensitivity of 52.1% with 1.89 false positives per image. No external generalization study was performed.

Although ANN or CNN algorithms are typically used as a means of false positive reduction, [5] showed that support vector machine (SVM) was also a suitable machine learning classifier for this task. Their best SVM models achieved a sensitivity of 71% at 1.5 false positives per image, 78% at 2.5 false positives per image, 85% at 4 false positives per image. At the highest achieved sensitivity of 92% and 100% false positives per image were 7 and 8, respectively. These results were achieved using a pre-processing pipeline lung field segmentation, histogram equalization, and difference imaging. SVM alone resulted in 70% sensitivity at 16 false positives per image, which was unsatisfactory, therefore the authors calculated a set of 160 features using techniques such as shape/size analysis, grey level distribution, and multi-scale LoG and applied a univariate Golub statistic to tune the scheme to achieve the reported results.

A common theme of the above studies is the use of a very large feature vector, with feature extraction accomplished by methods such as shape, intensity, texture, and gradient analysis. Unfortunately, where datasets are small, as is the case with lung nodule datasets including JSRT, with only 154 nodule positive examples, the number of features can be close to, or even larger than the number of discrete nodules resulting in a lack of robustness since the developed algorithms are easily overfitted to the small data corpus. This problem was called out by [13] who instead proposed augmenting nineteen texture/gradient-based features with a feature describing the degree of nodule isolation, since early-stage lung cancer may be characterized by the presence of a solitary pulmonary nodule. AdaBoost [52] was used to extract a white-nodule likeness feature map. This scheme resulted in an algorithm with a sensitivity of 80% at two false positives per image and a sensitivity of 93% at 5 false positives per image against the JSRT dataset.

The leading study into the use of deep learning for the lung nodule segmentation task is [21]. This study used sliding-window multi-resolution training of three different resolution CNNs with fully fused outputs. This method achieved the current state-of-the-art result against the JSRT-A dataset with a sensitivity of 90% at a false positive rate of 0.2 per image and a sensitivity of 100% achieved with a false positive rate of 0.4 per image. The results reported in this study are much improved than other studies in the literature and, in addition, are significantly better than the JSRT human expert radiologist metrics reported in the source paper [23]. Testing against two external datasets yielded a sensitivity of 45% and 50%, respectively. Given the very impressive internal testing result, external testing showed a disappointing generalization capability which was addressed in the study by fine-tuning the CNN to the external datasets.

A summary of works on automated lung nodule segmentation is shown in Table 1 along with the technique used for feature extraction and classification, preprocessing methods, and results of external testing where available.

**Table 1.** Summary of works relating to automated lung nodule segmentation using the JSRT dataset.

| Reference | Training Dataset | External Dataset | Image Size (pixels) | Pre-processing | Feature Extraction | Classifier | Internal Sensitivity (%) | Internal FP Rate | External Sensitivity (%) | External FP Rate | Notes |
|---|---|---|---|---|---|---|---|---|---|---|---|
| **[15]** | JSRT Subset (N=119) | University of Florence private dataset. (N=40) | 128 x 128 | Gabor Filter Lung Field Segmentation | Multiscale LoG | ANN | 75.0 | 10.2 | 75 | 8.9 | Excluded nodules < 5mm and > 20 mm. |
| **[5]** | JSRT (all) (N=247) | 240 Niguarda Ca'Granda Hospital in Milan (N=240) | 256 x 256 | Gaussian filtered image differencing | Multiscale circle finding and LoG | SVM | 100.0 | 8.5 | 95 | N/A | Total of 160 calculated features. External testing false positive rate was not specified. |
| **[53]** | JSRT (N=154) | None | 256 x 256 | Gaussian filtered image differencing | Multiscale LoG Ray casting | kNN | 67.0 | 4 | N/A | N/A | Noted that lung field segmentation had no positive effect. |
| **[14]** | JSRT-A (N=140) | None | 512 x 512 | Local contrast enhancement Lung field segmentation | Convergence Index Filter | Fisher's Linear Discriminant (FLD) | 80.0 | 5 | N/A | N/A | Used an adjacency rule to eliminate false positives resulting in approx. 10% improvement to sensitivity at FP rate. |
| **[20]** | JSRT-A (N=140) | None | Not Specified | Lung field segmentation Morphological operations and | Clustering watershed. | SVM | 78.6 | 5 | N/A | N/A | Used an adjacency rule to eliminate false positives. Inference times of 70 seconds. |
| **[13]** | JSRT-A (N=140) | Beijing Capital Intl Airport Hospital (N=15) | 1024 x 1024 | Lung field segmentation Rib suppression | Wavelet Transform + Convergence Index Filter | SVM | 90.0 | 4 | 100 | 4 | Used an adjacency rule to eliminate false positives. |
| **[21]** | JSRT-A (N=140) | Guangzhou Hospital (GDH) (N=168) Shenzen Hospital (SZH)l (N=240) | **Patches:** 224 x 224 168 x 168 112 x 112 | Lung field segmentation Rib Suppression Histogram Equalization | Multi-Resolution patch-based CNN. | Late-fusion multi-resolution CNNs | 99.0 | < 0.2 | 45 (GDH) 50 (SZH) | 1 1 | Pixel-level classification is resource intensive. SOTA internal testing results, but ~55% decrease in AUC for external testing. |

.

A clinically useful system must be both efficient and well-generalized. Thus, the key objective of this study is to find an efficient computer vision method/architecture that can locate lung nodules in CXR images, with algorithm generalization considered a higher priority than internal fitting. The main contributions of this work are as follows:

1. Establishing for the first time that an end-to-end deep-learning-based E-D network can *generally* segment/localize lung nodules when trained on a small data corpus of full-resolution CXR images.
2. We show that the best results are achieved using full-resolution images as input images to the E-D network. This result calls into question the very common practice of down-sampling images to fit out-of-the-box CNN input dimensions. We show that addition of convolution layers to the E-D network is a a better strategy for coping with large images than down-sampling.
3. Systematically showing that the intuitive, but unproven, image pre-processing steps of lung field segmentation and histogram equalization drastically algorithm generalization.
4. In our assessment we take the novel step of investigating the limits of the proposed algorithm in relation to:
    a. Nodule subtlety rating,
    b. Nodule malignancy,
    c. Nodule location,
    d. Patient diagnosis, and
    e. Patient sex.

## 2 Method

**2.1 Study Overview**

The study makes use of the frequently cited JSRT-A dataset as detailed in section 2.6.1 as a training dataset for three implementations of a residual E-D-based segmentation algorithm [54] based on U-Net, initially described by [22] and commonly employed for biomedical segmentation tasks. Three different depths of E-D algorithms are tested at 5, 6 and 7 layers for each encoder/decoder branch. The training dataset is divided by preprocessing options using histogram equalization, lung field segmentation, and a combination of these.

Each algorithm and pre-processing option are tested using three CXR image resolutions 512 x 512 pixels, 1024 x 1024 pixels, and 2048 x 2048 pixels with the latter being the full resolution of the JSRT images, and typical of the resolution of DICOM images captured in a clinical context. Each training dataset is split into 10 equally sized groups to facilitate 10-fold cross-validation. Splits across training datasets for different pre-processing options and CXR resolutions are made up of identical images to ensure that observed performance differences are the result of the controlled testing variable rather than an artefact of dataset splitting. External testing is performed against a curated version of the NIH ChestX-ray14 dataset [55] since this dataset provides location and size metadata for a subset of nodule images. This sequence of experiments is illustrated in Figure 1.

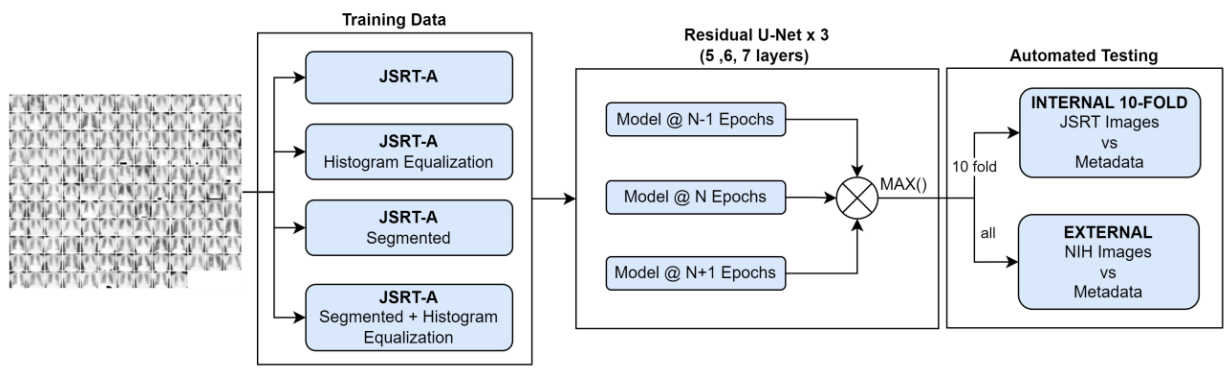

**Figure 1.** Summary of experiments performed in this study.

## 2.2 Model Composition and Training

Each E-D variation is trained up to 25 epochs at a learning rate of 1e-5 for each resolution and pre-processing option, with model weights captured every epoch. It was noticed during validation that lung nodules at different locations, and with different subtlety ratings, were located by the algorithm at different training epochs, offset from the point at which validation metrics stabilized before overfitting. Instead of there being a single optimal training epoch, the validation optimum is a saddle-point spread over three epochs. Each of the tested algorithms was therefore composed as an ensemble of three trained E-D models combined using simple element-wise maxima functions over the output arrays of each component model. The highest pixel value from each model output was taken as the final pixel value in a composite lung nodule prediction mask.

Figure 2 provides an example of an E-D training curve showing no consistent reduction in validation loss following epoch 12. This would tend to indicate that a model trained to Epoch 12 would be optimal.

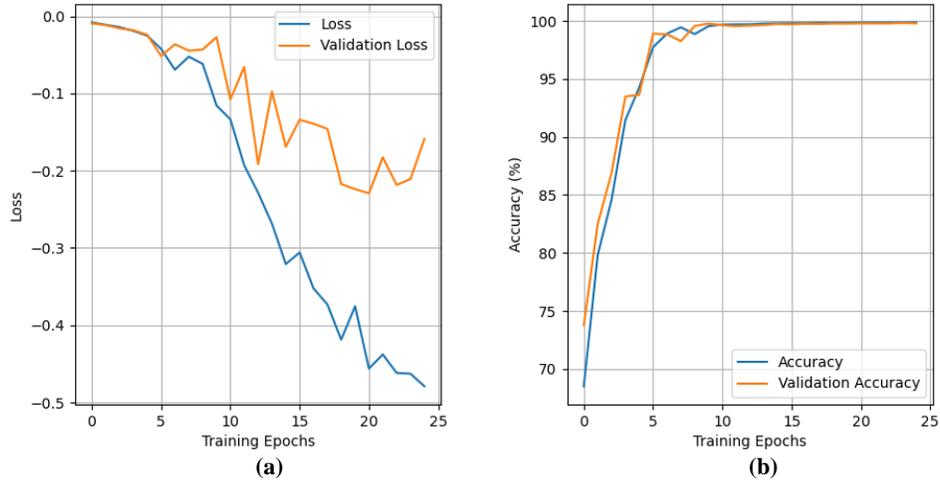

**Figure 2.** Training Curve example for E-D Model. Training Epochs vs Loss (a) and Training Epochs vs Accuracy (b).

For the obvious example of a lung nodule shown in Figure 3, epoch 12 is indeed the optimal model. In this example, the obvious nodule in file JPCLN005 is cleanly located with good segmentation at epoch 12, with epoch 11 being noisy, and epoch 13 still good, but starting to erode the nodule detection boundary.

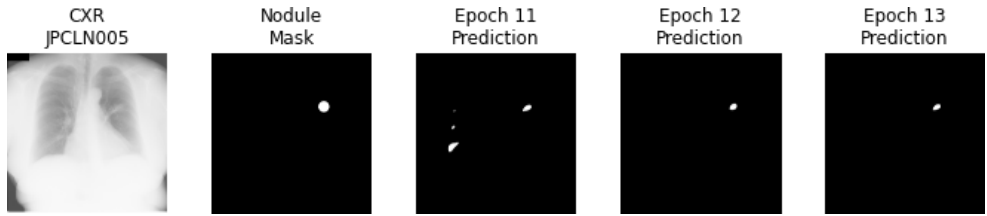

**Figure 3.** Example of obvious lung nodule correctly located at optimal training epoch.

However, more subtle nodules are located earlier in the model training. JPCLN146 is described as extremely subtle. As illustrated in Figure 4, this nodule is small, and completely missed by the algorithm at epoch 12 onwards. The nodule is, however, visibly segmented in the prediction at epoch 11, along with several false positives. Since false positives are preferable to false negatives, epoch 11 should be included in the overall predicted nodule mask for this image.

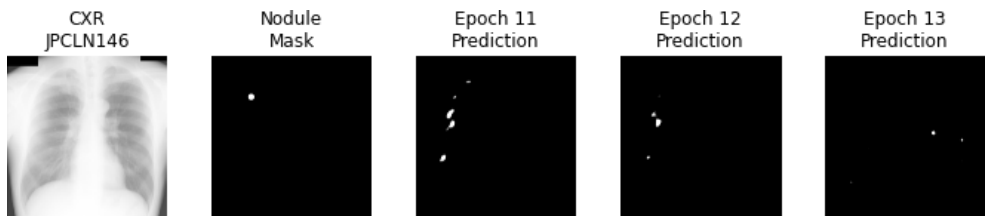

**Figure 4.** Extremely subtle lung nodules are located by earlier epoch in the model training.

Therefore, the proposed segmentation algorithm uses a simple combination of three trained models as described in Equation 1:

$$M' = max(M(n-1), M(n), M(n+1)) \tag{1}$$

Where $M'$ is the composition of model output arrays at optimal training epoch $n$, determined by training validation statistics. The output nodule mask prediction from $M'$ is therefore the element-wise maxima of predicted output arrays from the optimally trained model and the model trained to one epoch on either side of that trained model. Combining the algorithm output in this manner reinforces the true positives, with a relative de-emphasis of noise and false positives, resulting in cleaner nodule location prediction masks.

### 2.3 Automated Nodule Rating

Algorithm performance rating by manual visual comparison is time-consuming and inherently subjective. Instead, this study uses a fully automatic framework for nodule localization rating, using a blob detection algorithm derived from the OpenCV [56] contour finding function, followed by ellipse fitting to detect blob-like activations on predicted nodule masks. Fitted ellipses were checked for eccentricity being less than 0.95 and the area of the ellipse is greater than the total number of pixels in the mask (square of the dimension) divided by the square of tolerance of 128 pixels. For a full-sized image of 2048 pixels, this translates to a minimum area of 256 pixels, or approximately 3mm diameter anatomically, since the JSRT pixel size is 0.175 mm. Detected blobs outside of these parameters were considered noise artefacts from the decoding process and excluded from automated rating since the smallest nodule detectable on CXR images is around 10mm [57], and nodules found/confirmed using CT of size smaller than 6mm are considered to be a micro-nodules that do not require further investigation [58].

A region of interest (ROI) was then calculated using the ellipse center co-ordinates and the minor axis length. If the average pixel intensity within this ROI was less than 0.5, then the detection was filtered out since the ensemble output max function would reinforce a true positive lung nodule signal. These checks filtered out noise in the predicted nodule mask image, leaving suitable nodule candidates for checking against the JSRT metadata for centroid correlation. If the centroid co-ordinates of a fitted ellipse fell within a tolerance of 1/16[th] image dimension tolerance to the JSRT metadata nodule center location, then that detection was logged as a true positive. Other detections were logged as false positives. For full-resolution 2048 x 2048-pixel images, this translates to a 128-pixel tolerance, or approximately 22.4 mm anatomical centroid difference threshold for a detection to be considered a true positive, which is consistent with other studies using 20 – 25 mm tolerance [14, 21], but computationally neater for the purposes of a fair comparison of different image resolutions. This process is illustrated in the flowchart in figure 5.

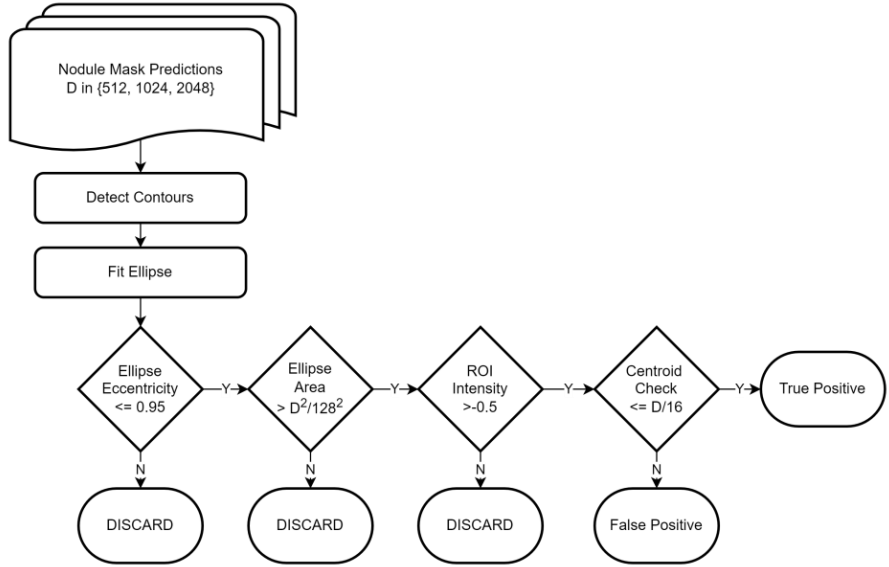

**Figure 5.** Automated nodule localization flowchart.

**2.4 Network Design**

Each experiment was performed using a residual U-Net, being an E-D network with a ResNet backbone [59]. We have previously used this architecture to match state-of-the-art results in the lung field segmentation task [60] as part of a pre-processing pipeline for COVID-19 severity assessment.

The largest input convolution filter in a standard U-Net E-D network is 256 x 256 pixels in size. This filter size is suitable for many applications, where a high-resolution image is either down-sampled to a lower resolution or divided into smaller patches prior to training. For full-resolution images of 2048 x 2048 pixels, with nodule features occupying only a small part of the image, a filter of this size could be expected to miss important features as the network builds an internal representational feature map. To address this problem, we modified the familiar residual U-Net architecture by implementing additional higher dimensional input and output convolution filters.

Three networks were coded using TensorFlow [61] at different depths consisting of 5, 6, and 7 layers designated E-D5, E-D6, and E-D7 with filter dimensions and several trainable parameters as detailed in Table 2. Note that E-D5 is a standard residual U-Net design, with E-D6 and E-D7 adding depth to the E-D network by further convolving the subject image, doubling the number of image channels, and halving the representational image dimensions at each additional layer. This has the impact of quadrupling the number of trainable parameters for each added layer in the network.

Table 2. Encoder-decoder network summary.

| Model | Network Depth | Network Filter Dimensions | Image Attributes at Full Encoding | Trainable Parameters |
|---|---|---|---|---|
| E-D5 | 5 | [16, 32, 64, 128, 256] | 256 x 32 x 32 | 4,715,441 |
| E-D6 | 6 | [16, 32, 64, 128, 256, 512] | 512 x 64 x 64 | 18,882,481 |
| E-D7 | 7 | [16, 32, 64, 128, 256, 512, 1024] | 1024 x 32 x 32 | 75,528,113 |

Figure 6 illustrates the structure of the deepest network used in this study E-D7.

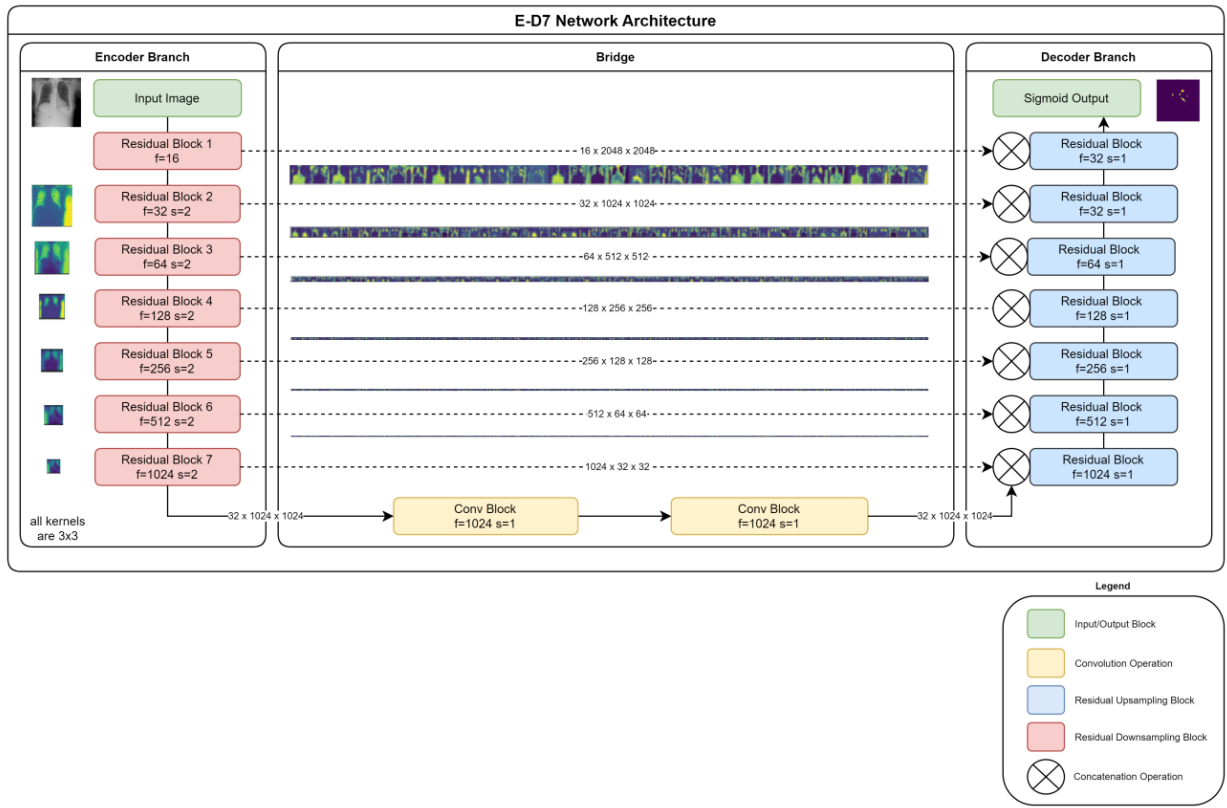

**Figure 6.** Schematic overview of the E-D7 Network.

## 2.5 Instance Normalization vs. Batch Normalization

Deep neural networks require normalization at the input and between convolutional layers to ensure that detected features are scaled for training network parameters. Without normalization, subtle features with smaller magnitudes would be swamped by obvious features with large magnitudes. Most CNN-based computer vision algorithms employ batch normalization, which trains neural network parameters normalizing using the mean and variance of a batch of samples. Since statistical mean and variance are used for batch normalization, anomalous signals caused by minority adversarial, or difficult, images in the batch can dramatically influence the normalization function and prevent the network from properly converging.

Instance normalization normalizes features over a single training image sample using a mean and variance that is calculated over the pixel array for that image [25]. Lung nodules appear as very weak signals on the CXR image, which are easily confused with rib crossings, vascular structures, and other thoracic anatomical features. The nodule feature only occupies only a tiny proportion of the overall image, which is quite different from the segmentation tasks for which U-Net is normally used. We found by experiment that nodule segmentation metrics using batch normalization were poor in comparison to instance normalization. Table 3 shows a comparison of internal test results at full image resolution between each E-D network with batch normalization versus instance normalization for the experiment using histogram equalization and lung field segmentation, with the normalization method as a controlled variable. Results from batch normalization were poor in comparison to instance normalization, and instance normalization has therefore been used for the remainder of the experiments in this study. This observation is probably a significant factor in the lack of any other study in the literature taking the E-D approach to CXR lung nodule segmentation.

**Table 3.** Comparison of instance normalization vs. batch normalization.

| Model | Instance Normalization | | Batch Normalization | |
| --- | --- | --- | --- | --- |
| | Sensitivity (%) | FP | Sensitivity (%) | FP |
| E-D7 | 81.4 | 6.6 | 2.9 | 0.5 |
| E-D6 | 85.0 | 7.9 | 37.9 | 6.4 |
| E-D5 | 61.4 | 8.4 | 45.0 | 9.5 |

The superior performance of instance normalization results from a combination of the subtle lung nodule signal along with an anatomical variation between CXR images. Statistical "smoothing" resulting from batch normalization would dilute an already weak signal. Instance normalization reduces this smoothing, and thereby allows the deep-learning model better to converge. This insight may have significance beyond the lung-nodule case in this study, to other datasets where the feature of interest presents a very low signal-to-noise ratio.

**2.6 Data Sets**

Two independent datasets are used in this study. Firstly, the JSRT dataset [23] providing a small, high-quality corpus of CXR images featuring solitary pulmonary nodules has been used as the training and internal testing dataset. External testing is performed against the NIH Chest X-ray14 dataset [55] providing a challenging, clinically realistic subset of CXR images featuring located nodules. Testing against this external dataset provides a good indication of the proposed algorithm's real-world usefulness.

**2.6.1     Training and Internal Testing Dataset**

Development studies showing the use of computer vision models for the automated identification of lung nodules from CXR images almost universally make use of the publicly available JSRT dataset [23]. This dataset comprises 93 control CXR images without lung nodules, and 154 CXR images featuring solitary lung nodules graded from obvious to extremely subtle. Each of the 154 solitary nodules CXR images are accompanied by metadata describing the age and gender demographics of the patient, the malignancy of the lung nodule along with the final diagnosis, and crucially for this study, the spatial co-ordinates and size of the lung nodule.

The JSRT image corpus is small for deep learning segmentation applications, which are normally trained on many hundreds to thousands of images [62] for the simpler task of organ segmentation. The solitary lung nodule only occupies a tiny portion of the overall CXR image, making nodule segmentation a more difficult task than organ segmentation where the organs of interest occupy a sizable part of the medical image, or biomedical cell segmentation where there are many cells of interest in each image for a U-Net to train on. Lack of disease-positive training data can pose a problem for lung nodule location studies, since ML and deep-learning systems tend to overfit small datasets, meaning that the deep neural network can effectively memorize the training dataset features with poor generalization to external datasets. For example, the leading nodule segmentation study achieved near-perfect nodule detection against the JSRT dataset under internal testing, however, an external generalization test yielded relatively disappointing cross-database performance with an average 55% decline compared to internal classification metrics [21].

To keep our study consistent with the literature, we have excluded from the JSRT nodule corpus 14 images where the identified nodule was outside the lung-field resulting in a total of 140 disease-positive images. This resulted in the so-called JSRT-A dataset as used by [13, 14, 21, 63, 64]. This exclusion is necessary for our experiment, since following segmentation, the generated lung nodule masks for these 14 images would be completely black – thereby invalidating comparisons between segmented experiments and unsegmented experiments. The set of JSRT-A images used in this study is presented in figure 7, showing (a) the original CXR images, (b) the effect of global histogram equalization, (c) lung-field segmented images, and (d) nodule mask for each image generated from JSRT metadata.

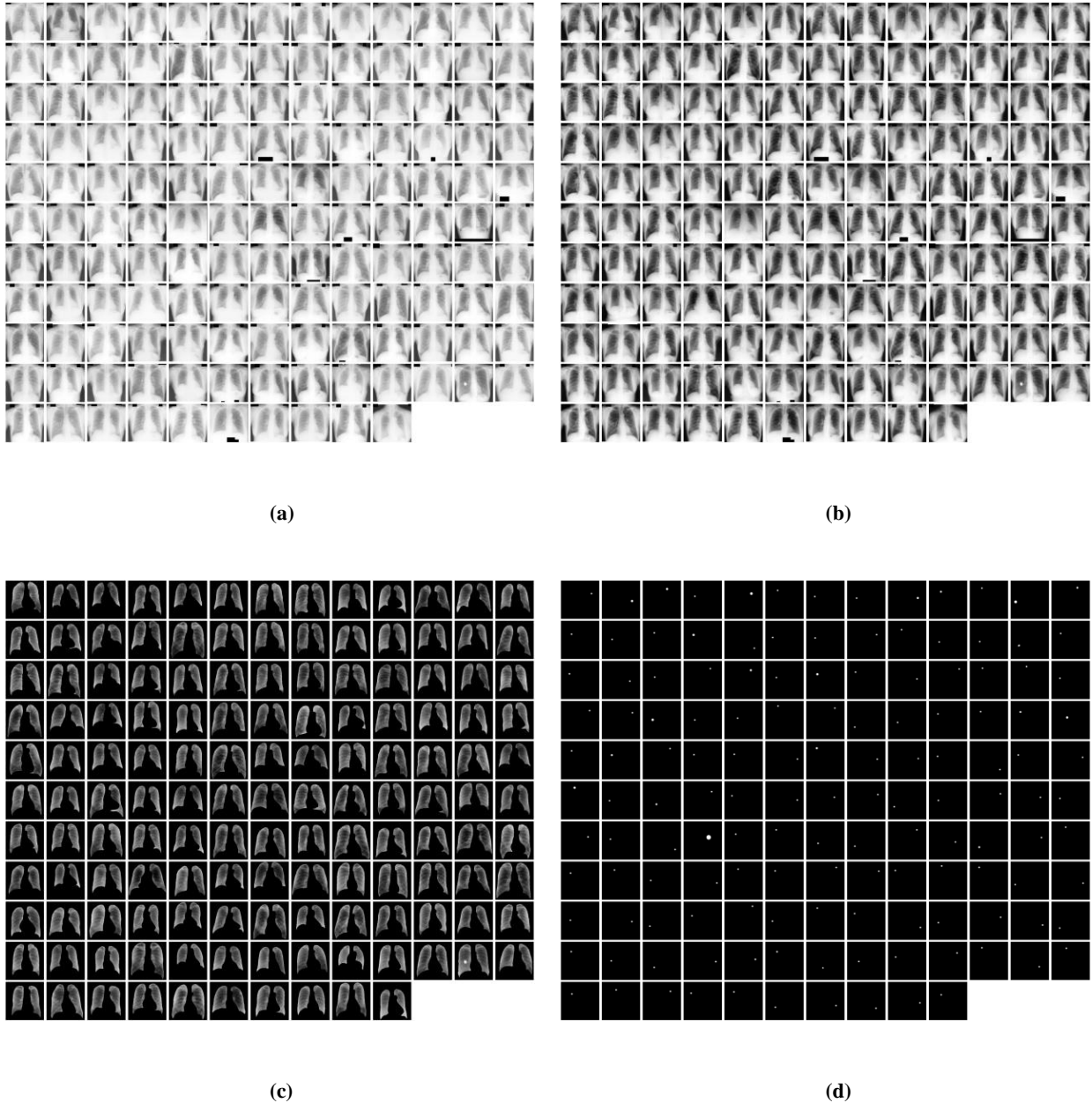

(a)          (b)

(c)          (d)

**Figure 7.** JSRT-A dataset summary montage showing: (a) base images; (b) the effect of histogram equalization; (c) the effect of lung-field segmentation using gold-standard masks; and (d) lung nodule locations and size using a circle approximation.

**2.6.2 External Testing Dataset**

A key element of this study is the objective determination of factors influencing computer vision model generalization to external datasets. A secondary CXR dataset was needed for this purpose, with the requirement that nodules had been identified and located on the CXR image with location metadata available. The publicly available NIH Chest X-ray14 dataset (NIH) [55] dataset consists of 112,120 frontal views CXR images of 30,805 unique patients, labelled with fourteen thoracic diseases that have been mined from accompanying radiology reports. CXR14 provides bounding box metadata of the labelled pathology for 984 of these images, of which 79 are classified as 'Nodule' and therefore suitable candidate images for this study. We noted that although the boundary box metadata singled out a single pulmonary nodule for the CXR image, several such images were heavily diseased and displayed more than one, and in some cases many, nodules along with multiple disease labels. These images were segmented poorly due to heavy disease obfuscation of the lung-field boundary and were discarded from the study. This resulted in 60 usable images for automated external generalization testing as shown in figure 8(a). Nodule masks were then generated for each of these images using the NIH boundary box metadata with the center of the circular nodule mask matching the center of the boundary box, and the diameter of the nodule mask calculated from the maxima of the width/height of the boundary box. This resulted in a set of NIH lung nodule masks as shown in figure 8(b).

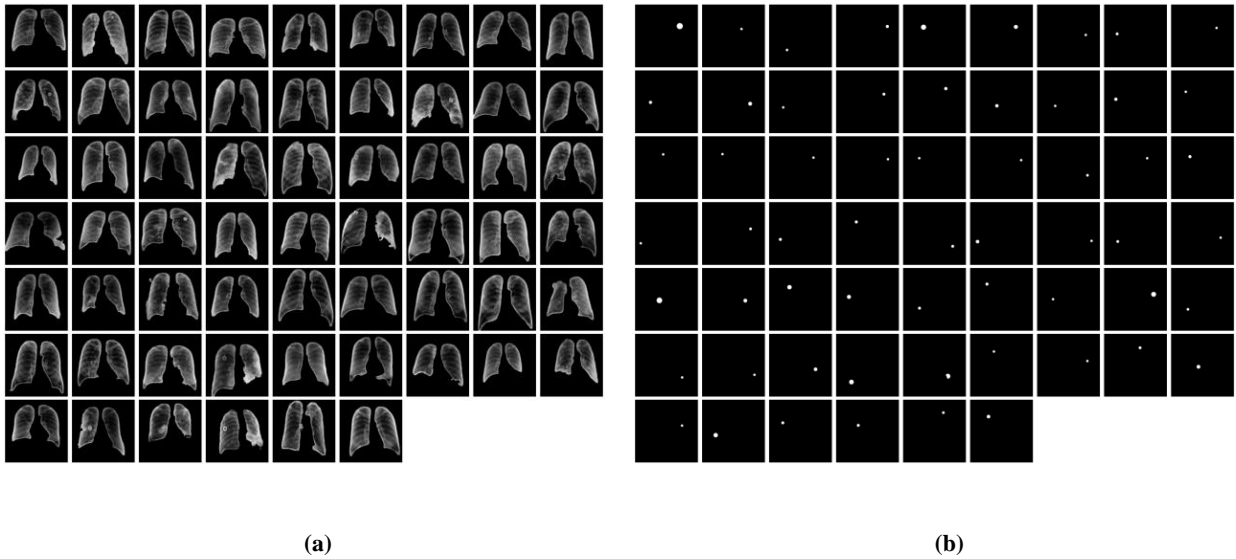

(a)          (b)

**Figure 8.** NIH image dataset summary montage showing: (a) segmented; and (b) lung nodule mask locations and size using a circle approximation.

## 2.7 Image Pre-Processing Operators

### 2.7.1 Histogram Equalization

Global histogram equalization (HE) is a contrast enhancement technique that is almost universally applied to medical images prior to computer vision analysis. HE is a simple transformer that distributes the most frequent pixel intensity values across an entire image. This operation globally expands the dynamic range of an image, intensifying pixel intensity differences. To the human eye, this makes contrast-based medical image features easier to distinguish [65-68], however, there are few studies into the effect of HE on computer vision algorithms. Our earlier work [69] concluded that HE promoted deep learning model generalization when combined with other techniques such as lung field segmentation and rib suppression. This improvement in generalization is because of HE removal of systematic differences in brightness and contrast between image acquisition sites, which would otherwise result in a model biased to the training site image acquisition parameters, leading to overly optimistic model performance metrics commonly reported from internal testing alone.

### 2.7.2 Lung-Field Segmentation

Since lung nodules appear in the lung-field, it stands to reason that the removal of any pixels not within the lung-field is a simple way to remove noise from the CXR image, allowing the computer vision algorithm to better learn lung nodule features. The JSRT dataset provides an accompanying set of gold-standard lung field masks that have been prepared by board-certified radiologists. These masks were used to create a lung-field segmented version of the JSRT-A dataset.

## 3 Results

### 3.1 Internal Testing at 512 x 512 pixels

Table 3 provides a summary of internal testing results at a resolution of 512 x 512 pixels. Only one test at 512 x 512 yielded sensitivity results above the 80% threshold, with the highest sensitivity of 81.4% at 7.3 false positive detections per image. The best results at this resolution were obtained using the shallowest 5-layer E-D which can be interpreted as undertraining of the deeper networks on the smaller image size. For this model, lung field segmentation and HE were effective in improving sensitivity and reducing false positive detections, with the combination of these methods yielding the best results, with a 10% improvement in sensitivity at 2.5 fewer false positives than training on the raw image set.

Table 4. Results from 10-fold internal testing of each ensemble at 512 x 512 pixels.

| Model | Raw Image | | Histogram Equalization | | Segmentation | | Histogram Equalization + Segmentation | |
|---|---|---|---|---|---|---|---|---|
| | Sensitivity (%) | FP | Sensitivity (%) | FP | Sensitivity (%) | FP | Sensitivity (%) | FP |
| E-D7 | 64.3 | 6.3 | 64.3 | 7.0 | 60.0 | 6.8 | 65.7 | 6.5 |
| E-D6 | 75.0 | 8.7 | 67.1 | 7.6 | 63.6 | 4.6 | 72.9 | 6.8 |
| E-D5 | 71.4 | 9.8 | 77.1 | 9.3 | 75.7 | 7.1 | 81.4 | 7.3 |

## 3.2 Internal Testing at 1024 x 1024 pixels

Table 5 provides a summary of internal testing results at a resolution of 1024 x 1024 pixels. At this resolution, two E-D network ensembles achieved sensitivity over the threshold of 80%, with the best result obtained from the 6-layer E-D6 network with a sensitivity of 83.8% at a false positive rate of 8.3 detections per image. The combination of lung field segmentation and HE improved sensitivity (by over 10%). However, this was at the expense of a small increase in FP detections contributed by the HE operator.

Table 5. Results from 10-fold internal testing of each ensemble at 1024 x 1024 pixels.

| Model | Raw Image | | Histogram Equalization | | Segmentation | | Histogram Equalization + Segmentation | |
|---|---|---|---|---|---|---|---|---|
| | Sensitivity (%) | FP | Sensitivity (%) | FP | Sensitivity (%) | FP | Sensitivity (%) | FP |
| E-D7 | 70.0 | 7.2 | 68.6 | 7.5 | 69.3 | 6.7 | 69.3 | 6.7 |
| E-D6 | 72.9 | 8.1 | 77.9 | 8.6 | 77.1 | 6.3 | 83.6 | 8.3 |
| E-D5 | 70.0 | 9.3 | 70.0 | 9.7 | 81.4 | 8.3 | 82.9 | 8.5 |

## 3.3 Internal Testing at 2048 x 2048 pixels (full resolution)

Table 6 provides a summary of internal testing results at a full resolution of 2048 x 2048 pixels. At this resolution two of the E-D network ensembles achieved sensitivity over the threshold of 80%. At full resolution, using HE and lung field segmentation, E-D6 achieved a sensitivity of 85% at 7.9 false positives per image. The deeper network, E-D7, also met the sensitivity criteria, with 81.4 at a false positive rate of 6.6. This result represents the lowest false positive rate for any of the experiments exceeding 80% sensitivity.

Table 6. Results from 10-fold internal testing of each ensemble at 2048 x 2048 pixels.

| Model | Raw Image | Histogram Equalization | Segmentation | Histogram Equalization + |

|  | Sensitivity (%) | FP | Sensitivity (%) | FP | Sensitivity (%) | FP | Segmentation Sensitivity (%) | FP |
|---|---|---|---|---|---|---|---|---|
| E-D7 | 65.0 | 7.9 | 78.6 | 8.4 | 75.0 | 6.2 | 81.4 | 6.6 |
| E-D6 | 69.3 | 9.6 | 72.9 | 8.7 | 82.1 | 7.4 | 85.0 | 7.9 |
| E-D5 | 57.9 | 9.1 | 61.4 | 9.7 | 78.6 | 10.2 | 61.4 | 8.4 |

## 3.4 Overall comparison of the Encoder-Decoder approach to JSRT radiologists

The results of a comparison between the best experiment at each input resolution and the JSRT Radiologist's performance is presented in table 7. The best E-D model sensitivity was comparable to, or better than the JSRT radiologists. However, apart from the obvious nodule case, the standard deviation across the 10-folds was higher than the human radiologists. This indicates that the sensitivity of the 10-fold models was influenced by training/testing data splits, with certain splits being more challenging than others.

At full resolution, the best-performing E-D model is more sensitive to subtle nodules at comparable standard deviation to the JSRT radiologists. For obvious and relatively obvious categories of nodules, lower-resolution E-D models outperformed the full-resolution model. This may be a result of the down-sampling "smoothing" effect noted by [14], assisting with the localization of the obvious categories of nodules. However, the same smoothing effect negatively impacts sensitivity to the subtle categories of nodules, resulting in overall better classification metrics using full-resolution images. This is an important result since the clinical utility of nodule detection algorithms would be to assist radiologists in the localization of subtle rather than obvious nodules.

**Table 7.** Breakdown of results from 10-fold internal testing with HE and segmentation compared with JSRT radiologists.

| Degree of Subtlety | JSRT Radiologists Sensitivity (%) | | 512 x 512-pixels Sensitivity (%) | | 1024 x 1024-pixels Sensitivity (%) | | 2048 x 2048-pixels Sensitivity (%) | |
|---|---|---|---|---|---|---|---|---|
| | Mean | SD | Mean | SD | Mean | SD | Mean | SD |
| Obvious | 99.58 | 1.86 | 100.00 | 0 | 100.00 | 0 | 91.67 | 28.87 |
| Relatively Obvious | 92.63 | 13.11 | 92.11 | 27.33 | 92.11 | 27.33 | 81.58 | 39.29 |
| Subtle | 75.70 | 22.35 | 83.33 | 37.66 | 87.50 | 33.42 | 91.67 | 27.93 |
| Very subtle | 54.66 | 23.98 | 78.26 | 42.17 | 65.22 | 48.70 | 73.91 | 44.90 |
| Extremely subtle | 29.60 | 15.90 | 47.37 | 51.30 | 68.42 | 47.76 | 57.89 | 50.73 |

## 3.5 Receiver operating characteristic for full-resolution input images

In comparison to more traditional feature extraction techniques such as LoG, the E-D architectures used in this study have few parameters that can be modified to trade-off sensitivity against false positive rates. Therefore, to better understand the limitations of the proposed algorithm this study employed a simple method of morphological opening followed by morphological closing using an elliptical kernel starting at 3 pixels and increasing to 90 pixels with increments of 3 pixels before rating output images from E-D6 and E-D7 at full resolution. An elliptical kernel was chosen over a rectangular kernel since detections are rounded rather than angular, matching the actual geometry of lung nodules. This resulted in the ROC curves in figures 9 (a) and (b), respectively. E-D6 shows a sensitivity of around 80% at a false positive rate of 7. E-D7 achieves a sensitivity of 80% at a false positive rate of around 6. Sensitivity for E-D7 falls off more rapidly than E-D6 so at a false positive rate of 3, E-D6 has a superior sensitivity of around 60% compared to E-D7 at 56%. This is a result of the deeper E-D7 providing cleaner and more accurate initial nodule localization, but with weaker detections that are more rapidly eroded by morphological operations. Nevertheless, E-D7 achieves an overall best result of 81% sensitivity at a false positive rate of 6.4 using a morphological operation kernel of size 6 pixels.

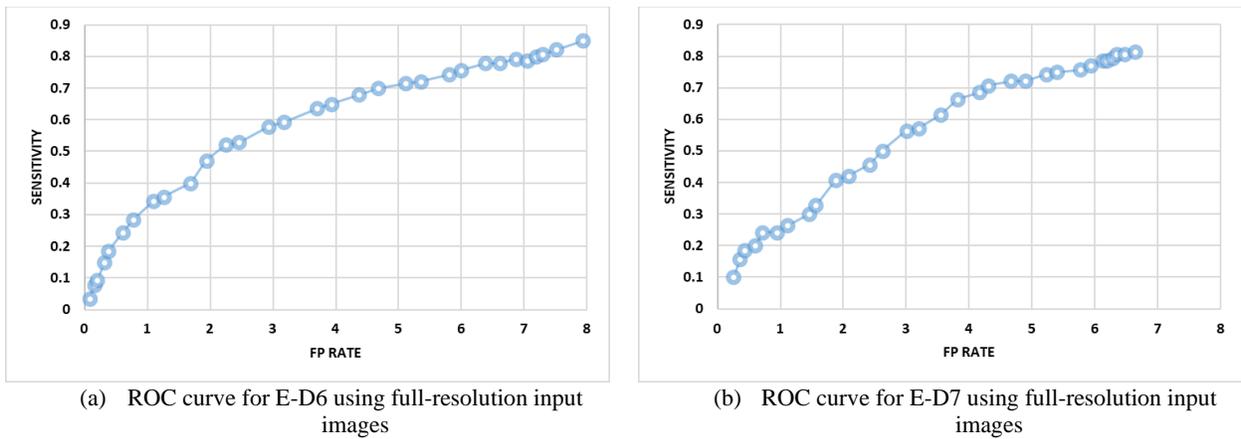

(a) ROC curve for E-D6 using full-resolution input images

(b) ROC curve for E-D7 using full-resolution input images

**Figure 9.** Comparison of ROC curves for best results at full resolution (2048 x 2048-pixel) input images.

### 3.6 Predicted nodule mask examples

Predicted nodule masks from 10-fold testing are presented in Figure 10, with a representative example from each subtlety rating.

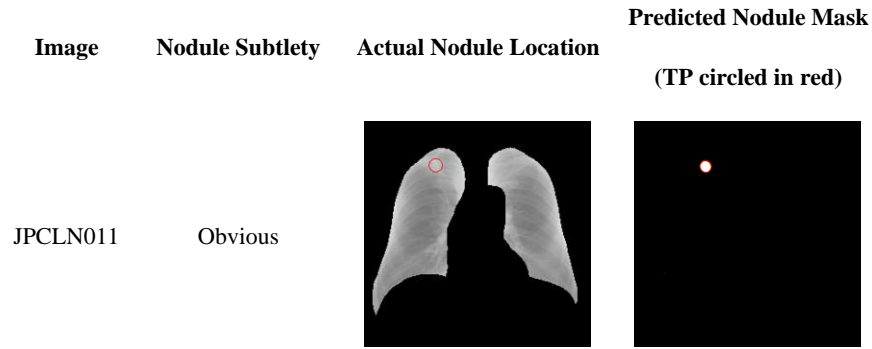

| Image | Nodule Subtlety | Actual Nodule Location | Predicted Nodule Mask (TP circled in red) |
|---|---|---|---|
| JPCLN011 | Obvious | | |

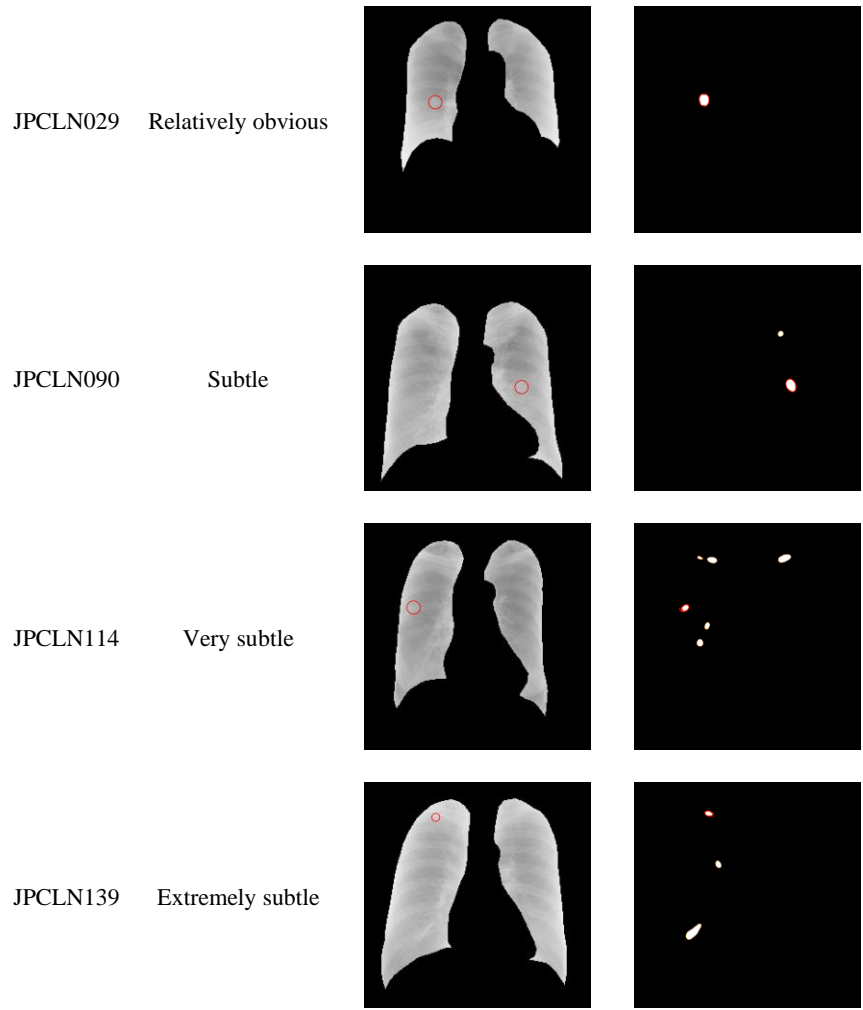

| | | | |
|---|---|---|---|
| JPCLN029 | Relatively obvious | | |
| JPCLN090 | Subtle | | |
| JPCLN114 | Very subtle | | |
| JPCLN139 | Extremely subtle | | |

**Figure 10.** Examples of nodule localization predictions from 10-fold testing.

The E-D model sensitivity to nodules is inverse to the subtlety of the nodules, and FP detections increase as the nodule subtlety increases. We noticed in inspecting the generated nodule masks that rib and collar-bone crossings were frequently the cause of false positive detections. Rib and bone suppression techniques such as dual energy CXR image acquisition could be expected to improve both sensitivities of this algorithm and reduce false positive rates.

### 3.7  Effect of variables on model sensitivity and FP rate

#### 3.7.1  Sensitivity and FP rate by nodule subtlety and malignancy

We were interested in whether the E-D architectures would be more sensitive to malignant versus benign nodules, potentially allowing this technique to rating the severity of detected nodules. Table 8 presents the results of E-D7 using HE and segmented full-resolution images separated by malignancy. The algorithm was more sensitive to obvious malignant nodules than obvious benign nodules, and conversely, was more sensitive to extremely subtle benign nodules than malignant benign nodules. This is a result of benign nodules tending to have a higher degree of calcification resulting in higher visibility on CXR images than subtle malignant nodules [70]. When averaged across 10 folds, sensitivity to benign nodules was 79.6% at 6.7 false positives per image, and sensitivity to malignant Nodules was 82.4% at 6.6 false positives per image.

Table 8. Results by the degree of subtlety and malignancy for E-D7 using HE and segmented images at full resolution.

| Degree of Subtlety | Malignant | | | Benign | | |
|---|---|---|---|---|---|---|
| | Sensitivity (%) | FP | N | Sensitivity (%) | FP | N |
| Obvious | 100.0 | 5.9 | 7 | 80.0 | 4.6 | 5 |
| Relatively Obvious | 85.7 | 6.8 | 28 | 70.0 | 7.0 | 10 |
| Subtle | 90.9 | 6.8 | 33 | 93.3 | 5.7 | 15 |
| Very subtle | 76.9 | 5.7 | 13 | 70.0 | 8.8 | 10 |
| Extremely subtle | 40.0 | 7.5 | 10 | 77.8 | 6.7 | 9 |
| **Mean** | **82.4** | **6.6** | **91** | **79.6** | **6.7** | **49** |

### 3.7.2 Sensitivity and false positive rate by nodule location

The JSRT dataset metadata describes the anatomical location of each nodule detection by side (left or right) and lung/lobe. The sensitivity of the E-D7 model using full-resolution images is mapped against these locations in figure 11. Sensitivity is relatively symmetrical across the left and right lobes, with the result for the right middle lobe being around 10% higher than for the left middle lobe. The most likely explanation for this result is that the heart and associated cardiological shadows are located on the left-hand side of the body, thereby obscuring the visibility of nodules making them more difficult for the algorithm to successfully detect.

Overall, the algorithm is much better at localizing nodules in the middle and lower lobes than in the upper lobes, regardless of whether the nodule is located on the left- or right-hand side.

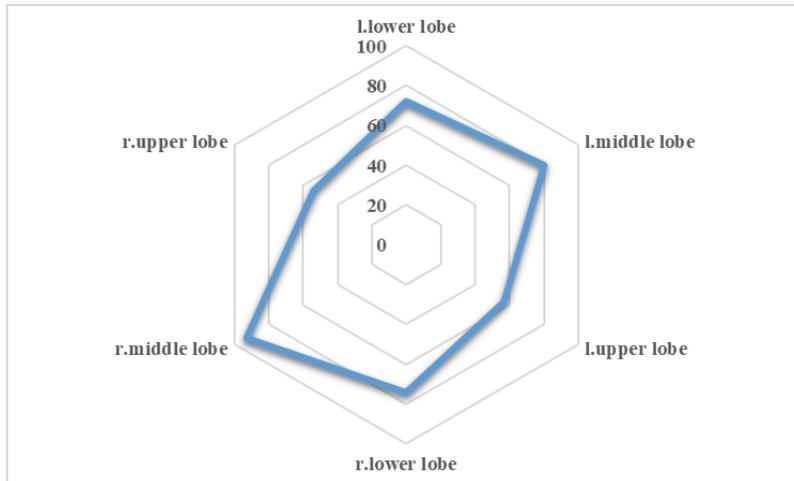

**Figure 11.** Sensitivity by nodule location.

### 3.7.3 Sensitivity and FP rate by sex

We were interested in the performance of the nodule localization algorithm by sex, before and after the applied image enhancements of HE and lung field segmentation. Sensitivity and FP rate results for each full-resolution test using E-D7 are presented in table 9. Initial results using raw images showed a significantly higher sensitivity for female over male CXR images, the application of HE and lung field segmentation significantly reduces this difference by debiasing the image samples.

**Table 9.** Sensitivity and FP rate by sex at full resolution using E-D7.

|  | Base Image | | Histogram Equalization | | Segmentation | | Histogram Equalization + Segmentation | |
|---|---|---|---|---|---|---|---|---|
|  | Sensitivity (%) | FP | Sensitivity (%) | FP | Sensitivity (%) | FP | Sensitivity (%) | FP |
| **Female** | 67.9 | 7.9 | 78.2 | 8.3 | 75.6 | 5.9 | 82.1 | 7.1 |
| **Male** | 61.3 | 8.1 | 79.0 | 8.4 | 74.2 | 6.4 | 80.6 | 6.0 |

### 3.7.4 Sensitivity and FP rate by diagnosis

Figure 12 shows the sensitivity of E-D7 using HE and segmented full-resolution images for each diagnosis in the JSRT Metadata. The algorithm is sensitive to most diagnoses at or above the 80% objective. There are nine diagnoses, excluding the "unknown" category, for which the algorithm provided lower than 80% sensitivity. Table 10 provides examples of these difficult images to highlight the limitations of the proposed algorithm. Notably, most of the missed detections relate to subtle categories of nodules that are obfuscated by ribs or are examples of conditions, such as pneumonia and cryptococcosis, which are very dissimilar to typical solitary pulmonary nodules associated with lung cancer.

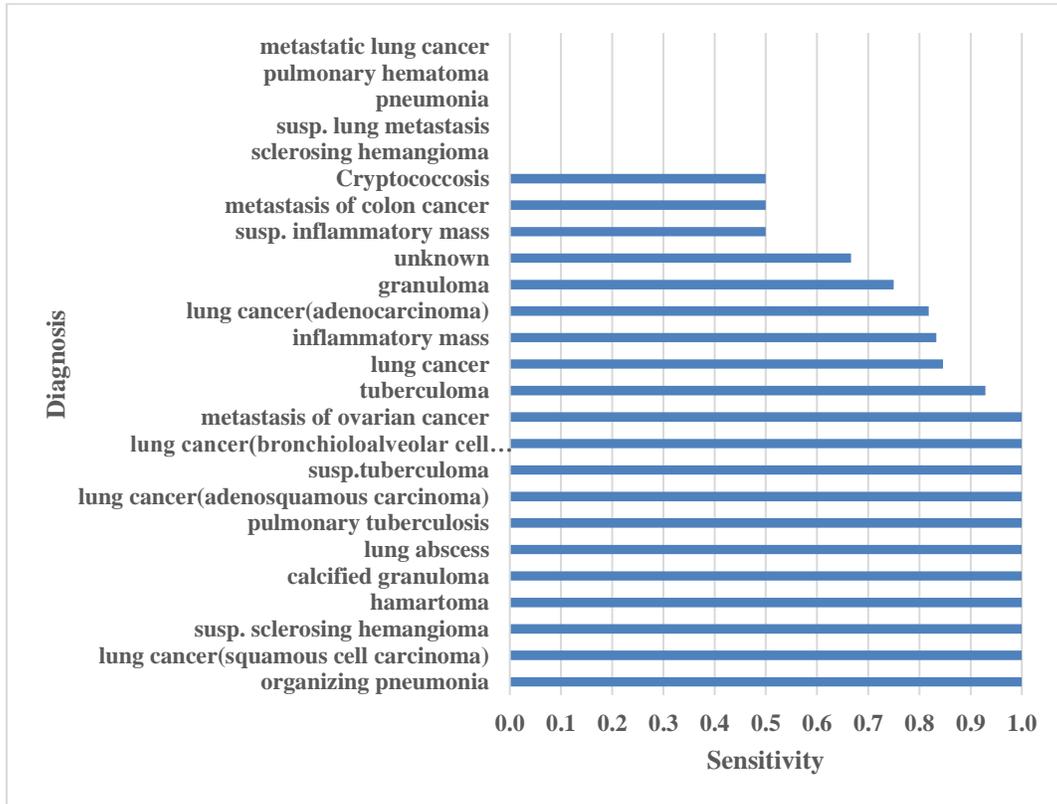

**Figure 12.** Sensitivity by diagnosis.

**Table 10.** Diagnoses/examples for which the E-D7 algorithm showed poor performance.

| Diagnosis | Sensitivity | Subtlety | N | Image Sample | Notes |
|---|---|---|---|---|---|
| Metastatic lung cancer | 0 | Extremely Subtle | 1 | 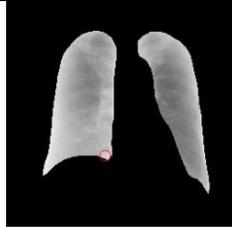 | Overlaps the inner edge of the right lower lobe. |
| Pulmonary hematoma | 0 | Obvious | 1 | 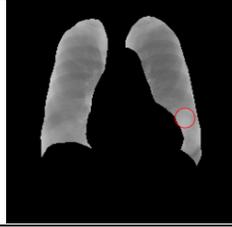 | Large blob occupying much of the lower right lobe. Little resemblance to a typical nodule in the training dataset. |
| Pneumonia | 0 | Very Subtle | 1 | 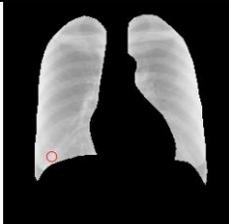 | Obfuscated by rib shadow at the lower extremity of the right lung. |
| Susp. Lung metastasis | 0 | Extremely Subtle | 1 | 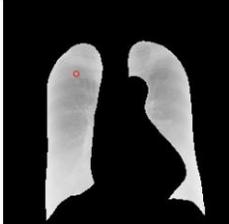 | Nodule located on the upper right lung. An extremely subtle nodule is obfuscated by a rib. |
| Sclerosing hemangioma | 0 | Extremely Subtle | 1 | 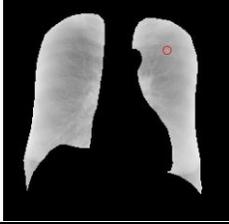 | An extremely subtle nodule is obfuscated by a rib crossing. |
| Cryptococcosis | 0.5 | Relatively Obvious | 2 | 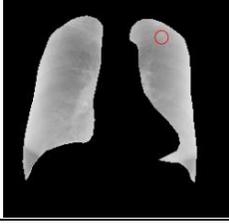 | Appears as an outline rather than a filled circle. Little resemblance to a typical nodule in the training dataset. |
| Metastasis of colon cancer | 0.5 | Subtle | 4 | 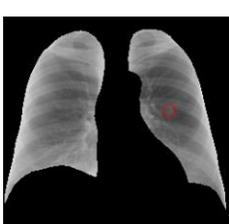 | Nodule appearance is very poorly defined and does not have the appearance of the solid circle. |
| Susp. inflammatory mass | 0.5 | Very Subtle | 2 | 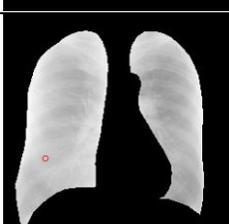 | Overexposed/low contrast part of the image. Very small dimensions for a suspected mass. |
| Granuloma | 0.75 | Subtle | 8 | 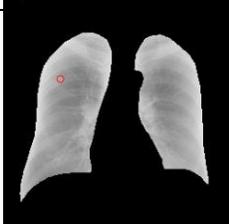 | A nodule is obfuscated by rib shadow. |

**3.8 External testing**

Other studies have also achieved excellent results against the JSRT dataset, sometimes with results surpassing label confidence by a large margin [21]. However, these studies typically neglect to subject JSRT-trained models to external testing. This study externally tested models trained on the JSRT-A dataset against the NIH dataset. Lung field segmentation and histogram equalization were used in all external tests since earlier work has indicated that these operations promote generalization [69].

**3.8.1    External testing for each model at various resolutions**

The first external test performed used models trained at each resolution using JSRT-A as a training corpus. The results of this test are presented in Table 11. An important consideration to note in interpreting these results is that, although the NIH metadata describes a single nodule per CXR image, many of the NIH images contain more than one nodule. The presence of multiple lung nodules on the NIH images has the effect of increasing the false positive count since some of the nominal false positives are true positives that have not been logged in the NIH metadata. Medical devices were another source of false positive detections, however, since such devices are pervasive in the clinical context, it is important to understand if the proposed algorithm was resilient to these, and if test images were not discarded based on the presence of medical devices. Illustrative examples are provided in figure 13.

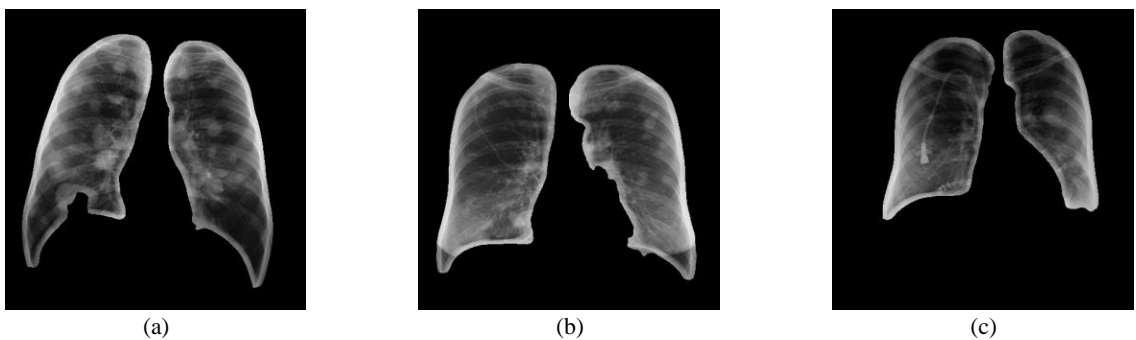

(a)        (b)        (c)

**Figure 13.** NIH Examples with (a, b) multiple pulmonary nodules; and (c) intrusive medical devices.

The best result was obtained at 1024 x 1024 pixels with a sensitivity of 78.3% at 13.2 false positives per image. Recalling that at 1024 x 1024 pixels internal 10-fold testing for JSRT-A using E-D5 achieved a sensitivity of 82.9% at 8.5 false positives per image, this represents a sensitivity penalty of only 4.6% with an increase in false positive rate of 4.6 per image. Bearing in mind that the false positive detections for NIH are artificially high due to the presence of multiple pulmonary nodules, we feel confident to conclude that the E-D5 variant of the proposed system has successfully generalized to the difficult and clinically realistic NIH dataset.

**Table 11.** External testing results.

| Model | 512 x 512 Sensitivity (%) | FP | 1024 x 1024 Sensitivity (%) | FP | 2048 x 2048 Sensitivity (%) | FP |
|---|---|---|---|---|---|---|
| **E-D7** | 50.0 | 16.4 | 60.0 | 9.9 | 70.0 | 13.4 |
| **E-D6** | 58.3 | 14.1 | 68.3 | 9.0 | 76.7 | 13.3 |
| **E-D5** | 78.3 | 16.1 | 78.3 | 13.2 | 78.3 | 15.0 |

Across these experiments, the best-performing E-D architecture is the shallowest 5-layer version. This is most likely a result of the deeper E-D6 and E-D7 networks having learned features of the JSRT training dataset that are specific to the JSRT dataset, resulting in network memorization and overfitting. The E-D6 and E-D7 networks would likely perform better if a larger and more diverse training corpus were available, such as could be obtained using a federated implementation. Upscaling the 1024 x 1024-pixel NIH images to 2048 x 2048 pixels did not improve sensitivity and increased false positives (due to noise introduced by interpolation) by 1.8 per image when using the E-D5 model.

Lung nodule localization examples resulting from external testing against the NIH dataset are presented in figure 14. Predictions made by E-D7 (c) provide a very clean segmentation of the metadata located in nodule (b), although missing some of the less obvious nodules that are apparent in the source image (a). These are better detected by the shallower networks E-D6 (d), and E-D5 (e). Figure 15(f) shows an example of an intrusive medical device on the right lung. This medical device proved to be a source of false positive detections, as all E-D models have incorrectly segmented this device as a nodule candidate (h-j). In practice, a human observer would easily dismiss these detections as obvious false positives, highlighting the importance of an expert overview of medical vision AI algorithms.

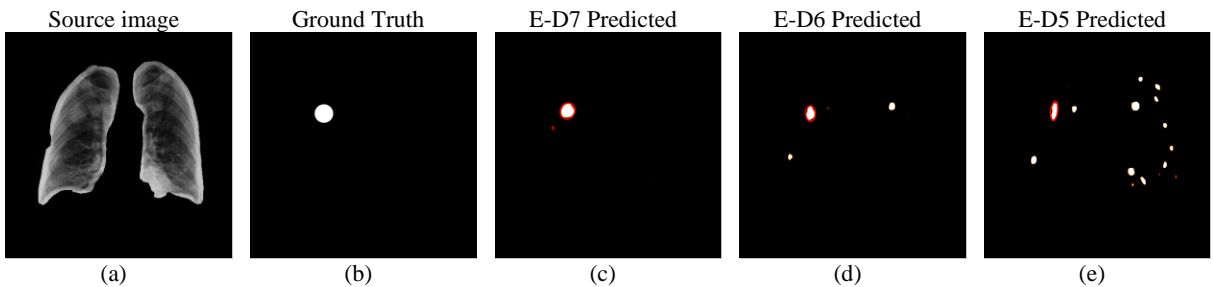

| Source image | Ground Truth | E-D7 Predicted | E-D6 Predicted | E-D5 Predicted |
| (a) | (b) | (c) | (d) | (e) |

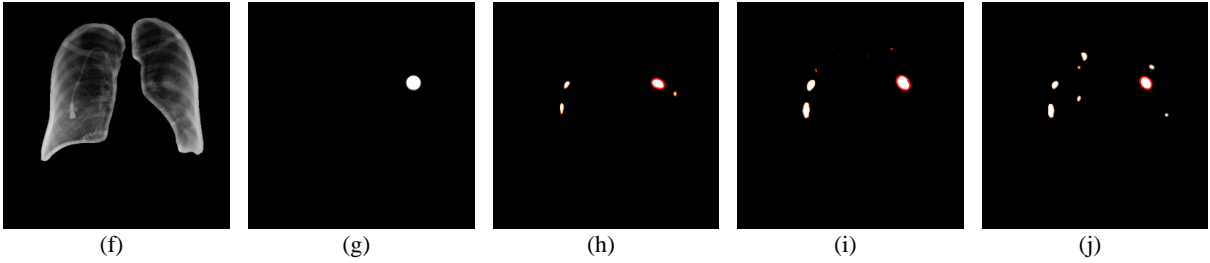

| (f) | (g) | (h) | (i) | (j) |

**Figure 14.** External testing using the NIH dataset examples showing: (c-e) sensitivity to nodules and false positive detections due to medical devices.

### 3.8.2 External testing with inclusions/exclusions of nodule subtlety categories

Noting that internal testing performance for the "extremely-subtle" and "very-subtle" categories of nodules were relatively weak, we sought to understand which subtlety categories should be included in model training to facilitate the best external testing results. We hypothesized that nodule masks for the "very-subtle" and "extremely-subtle" categories resulted in nodule masks that were little more than pinpoints. Along with the documented high inter-observer variability/disagreement in the data source metadata, this let us to suspect that these categories would have little bearing on the model performance and may even be detrimental using dilution of model learning. The JSRT-A training dataset was therefore split into four categories A through D by subtlety as described in table 12. E-D models were then trained on each of these splits, resulting in four models at each resolution trained on all nodule images through to only obvious nodule images. These models were then tested against the NIH dataset to establish the contribution of subtle categories of nodules to generalized model sensitivity.

**Table 12.** External testing categories definition.

| Category | Obvious | Rel. Obvious | Subtle | V. Subtle | E. Subtle | Samples (N) |
|---|---|---|---|---|---|---|
| A | ✓ | ✓ | ✓ | ✓ | ✓ | 140 |
| B | ✓ | ✓ | ✓ | ✓ | ✗ | 121 |
| C | ✓ | ✓ | ✓ | ✗ | ✗ | 98 |
| D | ✓ | ✓ | ✗ | ✗ | ✗ | 50 |

Table 13 provides the results of this test at a resolution of 512 x 512 pixels. The best result obtained at this resolution used the E-D6 model with very subtle and extremely subtle categories of nodules excluded. Best results were obtained for category C, excluding subtle and extremely subtle nodules, and using E-D6.

**Table 13.** Nodule categories external testing.

| Model | A | | B | | C | | D | |
|---|---|---|---|---|---|---|---|---|
| | Sensitivity | FP | Sensitivity | FP | Sensitivity | FP | Sensitivity | FP |

|  | (%) |  | (%) |  | (%) |  | (%) |  |
|---|---|---|---|---|---|---|---|---|
| E-D7 | 50.0 | 16.4 | 51.7 | 12.6 | 46.7 | 11.9 | 45.0 | 8.5 |
| E-D6 | 58.3 | 14.1 | 58.3 | 10.9 | 78.3 | 14.4 | 63.3 | 9.0 |
| E-D5 | 78.3 | 16.1 | 71.7 | 10.7 | 75.0 | 9.7 | 70.0 | 12.5 |

Table 14 shows the results of this testing at the NIH native resolution of 1024 x 1024 pixels. Best results were achieved using E-D5 with all categories of nodules included in the training data (A). Notably, excluding extremely subtle nodules per category B had little bearing on these results indicating that these images did not provide a significant training signal. Moreover, when the other E-D architectures are considered, the exclusion of very-subtle and extremely subtle nodules per category C led to a good result for E-D6 with a sensitivity of 76.7 with FP of 7.6, with this being the lowest FP value achieved for any external test. This result corresponds to the good result at 512 x 512 pixels for E-D6 with these categories of nodules excluded.

**Table 14.** Nodule categories external testing at 1024 x 1024 pixels.

| Model | A | | B | | C | | D | |
|---|---|---|---|---|---|---|---|---|
|  | Sensitivity (%) | FP | Sensitivity (%) | FP | Sensitivity (%) | FP | Sensitivity (%) | FP |
| E-D7 | 60.0 | 9.9 | 65.0 | 11.7 | 61.7 | 14.7 | 60.0 | 13.7 |
| E-D6 | 68.3 | 9.0 | 71.7 | 6.7 | 76.7 | 7.6 | 73.3 | 11.0 |
| E-D5 | 78.3 | 13.2 | 78.3 | 13.4 | 71.7 | 10.1 | 71.7 | 11.2 |

Finally, table 15 presents the result of this round of testing at 2048 x 2048 pixels. Once again, the best result was obtained using category C data, which excludes the very subtle and extremely subtle categories of nodules confirming that the inclusion of these categories of nodules in model training has a negative impact on external generalization.

**Table 15.** Nodule categories external testing at 2048 x 2048 pixels.

| Model | A | | B | | C | | D | |
|---|---|---|---|---|---|---|---|---|
|  | Sensitivity (%) | FP | Sensitivity (%) | FP | Sensitivity (%) | FP | Sensitivity (%) | FP |
| E-D7 | 70.0 | 13.4 | 71.7 | 10.6 | 71.7 | 8.8 | 66.7 | 8.3 |
| E-D6 | 76.7 | 13.3 | 75.0 | 14.9 | 70.0 | 16.1 | 78.3 | 12.6 |
| E-D5 | 78.3 | 15.0 | 76.7 | 16.5 | 85.0 | 16.4 | 78.3 | 14.3 |

## 4 Discussion

Many development studies present computer vision algorithms capable of diagnosing up to 14 different thoracic diseases from normal lungs, including lung nodules. Most of these studies solve the problem in either binary or multiclass form – giving a computer vision-based probability of a certain disease being present in the lung including the original NIH paper [55] and follow-up studies, with state-of-the-art methods achieving multi-class accuracy for nodule detection of 77.7% [71]. These studies normally use saliency techniques to display network attention to locate the disease of interest in sample images. Saliency techniques tend to have a relatively poor definition, whilst an E-D architecture can generate an image mask that cleanly segments the lung nodule, as shown in figure 15. Each of these masks was generated at a resolution of 1024 x 1024 pixels, with training samples excluding very subtle and extremely subtle nodule categories of JSRT images. This combination resulted in the cleanest mask images of all the experiments, with near pin-point nodule segmentation in external testing.

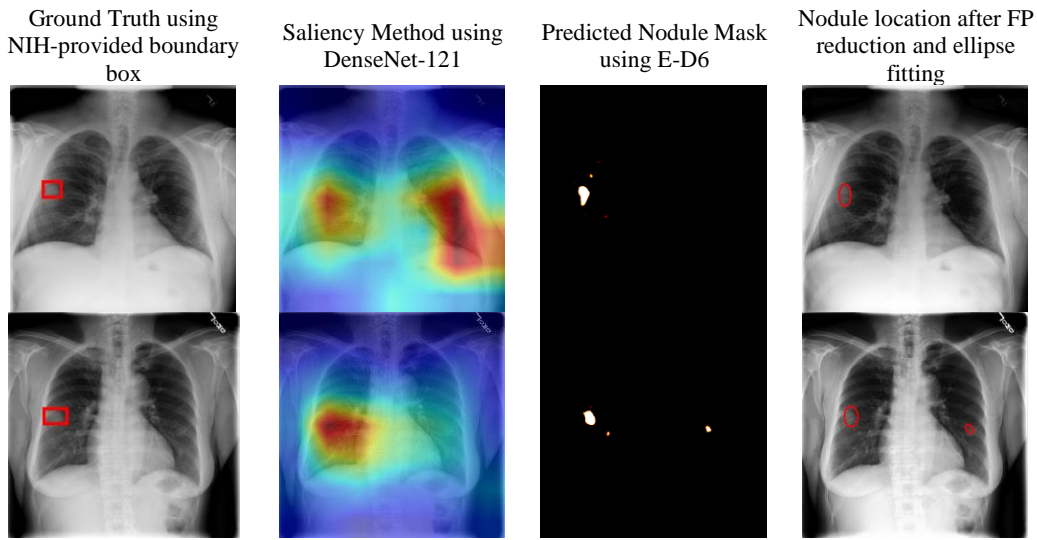

**Figure 15.** Example of a nodule located using saliency technique compared to the E-D method.

Other studies have successfully located lung nodules in the JSRT dataset, with good results. However, these studies have performed relatively poorly under external testing in the absence of fine-tuning to the external dataset [21]. Where external testing is performed, the external datasets used are typically a private set of images without indication of clinical realism and/or curation processes, making replication impossible [5, 13, 15, 21]. The tendency for AIs to worsen when applied to patient groups with different characteristics to model development groups is known as "data set shift" and has recently been called out as a major problem by [72], who propose evaluating an AI's effectiveness based on real-world clinical evaluation rather than stand-alone model performance, noting that this narrow focus has led to limited penetration of radiologic AI applications in practice.

Although our internal testing results are not as good as those [21], who exceed label confidence by a considerable margin, the proposed method is fast to train, at approximately 20 minutes per epoch at full resolution resulting in an overall training time of around 8 hours on the modest hardware used. Inference time for a full-resolution CXR measuring 2048 x 2048 pixels is 768ms, and 198ms at the NIH resolution of 1024 x 1024 pixels, which would easily fit into an augmented clinical workflow. This is much faster than inference times achieved by studies using complex linear filtering methods based around LoG, where inference times of up to 70s per image have been reported [20]. The output of the proposed algorithm is a nodule mask that could be overlaid on the source CXR image to produce an instantaneous visual cue to human radiologists to pay close attention to parts of the CXR image containing nodules.

This study has found that an E-D deep-learning-based architecture can effectively localize lung nodules when trained on a small data corpus of full-resolution CXR images. Accurate lung nodule segmentation is greatly promoted by substituting instance normalization for batch normalization at the E-D network layer boundaries. Instance normalization is not typically used in CXR deep learning applications, our use is novel, and this may explain why the use of E-D architectures such as U-Net has not been previously reported for lung nodule segmentation tasks. In our testing, good internal testing results did translate well to external testing with only a minimal drop in performance.

Higher resolution images yielded better results under internal testing conditions with 2048 x 2048-pixel images allowing for a top sensitivity of 89% with 7.9 false positive detections per image. The use of HE and lung-field segmentation had a significant positive effect on sensitivity and false positive rate under 10-fold testing. This result was found to be largely due to an improvement in sensitivity to the subtle categories of lung nodules. This result calls into question the standard practice of down-sampling CXR images to CNN default image size 299 x 299 pixels, or less, prior to running training. It is apparent from our testing that important pathological signals are weakened by the down-sampling process. This conclusion is supported by external testing achieving the best results at a resolution of 1024 x 1024 pixels, which is the native resolution of the external NIH dataset. Upscaling NIH images to 2048 x 2048 pixels did not improve sensitivity, and the interpolation noise resulting from upscaling caused an increase in the false positive rate for this test.

Reducing a 2048 x 2048-pixel JSRT image to 299 x 299 pixels would reduce the axial dimensions of a nodule 7-fold, and therefore, the area of the nodule by a factor of 49. At the JSRT pixel size of 0.175mm, this would reduce a 10mm diameter nodule to a 1.4mm feature on the 299 x 299-pixel image representing around a diameter of only 8 pixels. This is a very small feature to expect a deep learning neural network to train on, especially on a noisy image such as a CXR image. It follows that lung nodule computer vision algorithms should be trained on the highest available resolution of source CXR image within the processing capabilities of available compute resources.

Under internal testing conditions, the best results were achieved by 6 and 7-layer E-D networks. However, external testing results were best with a 5-layer network. Since deeper neural networks tend to overfit, this result may point to features in the JSRT dataset that the deeper networks can internally represent, that do not transfer to the NIH external dataset. We have shown that there exists a trade-off between network depth and generalization capability that should be tuned by the application based on the understating that internal-only testing can provide overly optimistic results for deeper networks.

It is important to understand the limitation of any tool, especially one that may be applied to human medical diagnosis. AI tools used for CXR computer vision will be used under the supervision of expert radiologists, who will easily note Obvious and Relatively Obvious categories of nodules without automated assistance. The study found that the JSRT expert radiologists outperformed the proposed algorithm for these categories. The situation is different for subtle categories of nodules, where the algorithm outperformed the JSRT radiologists on the sensitivity measure. These results indicate that the clinical usefulness of the algorithm in a clinical setting would be to draw radiologist attention to potentially missed subtle to extremely subtle nodules. One very important limitation to this use case is relatively poor sensitivity to malignant extremely subtle nodules, which was found to be 40%, compared to 78% for benign extremely subtle nodules. Unfortunately, these are the most desirable nodules to find early – since they are likely to grow into life-threatening cancers. We also noted that the algorithm was more sensitive to nodules in the middle and lower lung lobes than the upper lung lobes, most likely because of the presence of scapular and rib crossings mimicking nodules in these geographies. The team is currently working on a full-resolution rib and bone suppression system which is expected to mitigate this limitation in the future. We also noted that the algorithm was un-biased to sex, but only after the HE and lung field segmentation steps were applied.

In external testing, E-D6 provided an excellent generalization result when very subtle and extremely subtle nodules were excluded from the JSRT training data. Internally, at 1024 x 1024 pixels, E-D6 achieved a sensitivity of 83.6% with 8.3 false positives per image. Externally E-D6 achieved a sensitivity of 76.7% at 7.6 false positives per image. This result shows that the E-D6 algorithm can generalize well across datasets from different clinical origins. This is a promising result to create a federated implementation whereby separate clinics can contribute to a centralized model using local data, whilst preserving patient confidentiality – since only network weights are shared beyond the wall of the clinic. Ideally, several independent clinics could contribute to a centralized model that would learn sensitivity to more subtle nodules via access to a globally diverse and large training data corpus.

In comparing the performance of the full-resolution deep-learning approach against the performance of human radiologists using provided metadata, we found that the method presented is approximately equivalent in sensitivity to human radiologists for obvious nodules, but outperforms human radiologists for subtle, and very subtle nodules, although with a statistical deviation that is too wide to immediately support a clinical system. With more data, the deviation gap could be reduced, resulting in a fully generalized tool that could help improve the speed and quality of lung cancer diagnosis from CXR images, and thereby allow for earlier intervention and a better prognosis for sufferers of this deadly cancer.

## 5   Conclusion

This study proves that it is feasible to create a generalized lung nodule segmentation/localization algorithm using well-known E-D algorithms, albeit with several novel refinements. The results achieved are very promising, especially given the small size of the training dataset for a deep learning algorithm. The proposed algorithm does not rely upon any parameterized linear filtering methods such as LoG, instead the method used end-to-end deep-learning providing simultaneous feature extraction and semantic segmentation. This has the great advantage of allowing the algorithm to generalize to external datasets without re-tuning parameters or fine-tuning against the external dataset.

The motivation for this study was to establish the foundation for a clinically useful algorithm that could augment a radiologist's workflow. E-D networks such as those tested here would seem to be an ideal candidate for this purpose since they are resource efficient, implemented using standard libraries, and easily deployed as federated algorithms.

The main limitation of the proposed method is the availability of labelled and segmented training data for the E-D network. More data could be expected to improve segmentation accuracy leading to higher sensitivity and lower FP scores.

In the future, we will investigate two strategies for accessing additional data. Firstly, we will use the algorithm described in the paper to "bootstrap" training data from the NIH dataset, for external testing against a third independent dataset, with human radiologists providing sensitivity and FP scores. Secondly, we will implement the E-D algorithm as a swarm/federated learning [73, 74] model. When implemented as a federated clinical trial, the higher volume of more diverse CXR images would result in a tool that meets the needs of a commercial clinical system, whilst completely preserving patient privacy. Furthermore, low resource requirements, particularly on inference, make E-D-based designs suitable to be embedded in mobile and/or augmented reality applications supporting telehealth and improving radiologist access to remote, and disadvantaged communities. It is hoped that AI-based assistance may enable more widespread lung cancer screening via fast and inexpensive CXR images resulting in earlier intervention and mortality reduction.

**Conflict of Interest Statement:** The authors have no affiliation with any organization with a direct or indirect financial interest in the subject matter discussed in the manuscript.

**Funding:** This research did not receive any specific grant from funding agencies in the public, commercial, or not-for-profit sectors.

# 6 ACKNOWLEDGMENTS

The first author would like to thank his employer IBM Australia for providing flexibility allowing for the performance of experimental work and preparation of this manuscript.